\documentclass[12pt]{article}
\title{Scale Dependent Intermittency and Conformal Invariance in Turbulence}
\author{G.A. Kuzmin \\
Institute of Thermophysics, 630090 Novosibirsk, Russia\\ E-mail:
kuzmin@itp.nsc.ru}
\begin{document}

\maketitle

\begin{abstract}
We present a conformal theory for intermittent scalar fields. As an example,
we consider the energy flux from large to small scales in the developed
turbulent flow. The conformal correlation functions are found in the
inertial range of scales. In the simplest case, the theory leads to the
log-normal model. Parameters of the model are expressed via integrals of the
conformal correlation functions. Non-Gaussian conformal correlation
functions of high order are studied.
\end{abstract}

\bigskip PACS-number: 47.27 \newpage

\section{Introduction}

The present paper studies the scale and conformal symmetry of intermittent
turbulent pulsations at very large Reynolds number. The scaling ideas were
introduced into the turbulence theory by Kolmogorov \cite{kol41}. Later
those ideas were fruitfully explored in the second-order transition theory
\cite{PP79}, in the quantum field theory and so on. The conformal symmetry
is a powerful tool that strengthen the predictions of the scale symmetry.
Its application to the turbulence theory is tangled by the scale dependent
intermittency that destroys the simple scaling scheme.

Kolmogorov \cite{kol41} defines the inertial range of scales for which he
made two major suppositions. First, that the correlation functions of
fluctuating fields are invariant to translations, rotations and scale
transformations. Second, that the mean dissipation energy is the only
dimensional parameter determining the statistics in that range. That theory
was refined in 1962 \cite{kol62} in order to take into account the effects
of intermittency. The main result of the latter theory is the log-normal
model that predict corrections to the simple scaling. In addition, a
modification of that theory was proposed. The principal fields of the
modified scaling theory was the ratios of the velocity differences. That
fields were proposed to be statistically invariant according to scale
transformations.

Scaling determines the correlation function up to unknown dimensionless
universal functions. In order to obtain more information from symmetry
groups, the conformal symmetry was proposed as the simplest extension of the
scale one. The conformal group includes the special transformation that
locally looks as the scale one. The conformal theories were applied to a
wide set of physical problems \cite{MS69}, \cite{pol70}. It is possible, in
that theories, to determine the three point correlation function up to
constants and to diminish the number of universal dimensionless arguments in
the correlations of higher order. The conformal symmetry is especially
informative in two dimensions where any analytic function of a complex
variable induces some conformal transformation \cite{BPZ84}. An application
of the conformal symmetry to the fluid turbulence was considered in \cite
{KP74}. The conformal theory that based on the method of the paper \cite
{BPZ84} was studied in \cite{pol93}. The present paper does not use the
concepts of the latter method.

We consider the conformal theory for intermittent field in three dimensions.
The starting point of the present theory is the Kolmogorov (1962) refined
scaling for the dimensionless ratios of fields. We study application of the
conformal symmetry to correlation functions of energy dissipation. The
result is logarithmical normal theory that relates correlations in space and
scale, the expression for the constants of the log-normal model and
expressions for correlations of higher order.

\section{Simple and intermittent scaling}

\subsection{Simple scaling}

The local structure of the developed turbulence has rotational and
translational invariance. One considers the structure functions --- the
correlations of the velocity difference $\mathbf{w}(\mathbf{r})\mathbf{=u}(
\mathbf{x+r})-\mathbf{u}(\mathbf{x}).$ It is supposed that the structure
functions are invariant to rotations and translations. In the simple
scaling, the statistical regime is invariant to the scale transformations
\[
\mathbf{x}\rightarrow K\mathbf{x},K>0,
\]
\begin{equation}
\mathbf{u}(\mathbf{x})\rightarrow K^{\Delta _u}\mathbf{u}(K\mathbf{x}),
\label{scale}
\end{equation}
where $\Delta _u$ is a number called the scale dimension of velocity. In the
Kolmogorov (1941) theory of incompressible fluid, $\Delta _u=-1/3$. The pair
structure function is
\[
\left\langle w_i(\mathbf{r})w_j(\mathbf{r})\right\rangle =A\varepsilon
^{2/3}r^{2/3}\left( 4\delta _{ij}-\frac{r_ir_j}{r^2}\right) ,
\]
$A$ is a constant.

\subsection{Refined scaling for the dimensionless scalar fields}

For simplicity, we consider the Kolmogorov modified theory for dimensionless
ratios of the smoothed scalar fields. A reformulation of scaling for the
vector fields see in \cite{kuz96a}. To be definite, let us consider the
density of the energy dissipation

\[
\varepsilon (\mathbf{x})=\frac \nu 2{\left( \frac{\partial u_i(\mathbf{x})}{
\partial x_j}+\frac{\partial u_j(\mathbf{x})}{\partial x_i}\right) ^2}.
\]

The dissipation smoothed over a sphere of radius $l$ is
\[
\varepsilon (\mathbf{x},l)=\frac 3{4\pi l^3}\int\limits_{r\leq l}\varepsilon
( \mathbf{x}+\mathbf{r})d^3r.
\]
This field is believed to model the energy flux from large scale to the
small ones in the inertial range. Our approach is valid for any intermittent
scale invariant scalar field.

The ratio of dissipation smoothed over two different scales $l_1,l_2$ is
\begin{equation}
\psi \left( \mathbf{x},l_1,l_2\right) =\frac{\varepsilon (\mathbf{x},l_1)}{
\varepsilon (\mathbf{x},l_2)}.  \label{ratio}
\end{equation}

The dimensionless scalar $\psi \left( \mathbf{x},l_1,l_2\right) $\ is the
field of the same kind as the Kolmogorov's ratios of the velocity
differences. The present paper deals with inertial range where direct action
of viscosity is negligible. According to the refined theory \cite{kol62},
the correlations of the field $\psi $ have to be invariant to translations,
rotations and to scale transformation. The scale transformation is
\[
\psi \left( \mathbf{x},l_1,l_2\right) \rightarrow \psi \left( K\mathbf{x}
,Kl_1,Kl_2\right),
\]
where $K>0$ is any numeric multiplier. We suppose that the dimensionless
fields have zeroth scaling dimension. Invariance to the scale transformation
determines the pair correlations of $\psi $ up to an universal scalar
function $\Psi $ of dimensionless variables
\[
\left\langle \psi \left( \mathbf{x},l_1,l_2\right) \psi \left( \mathbf{x+r}
,l_3,l_4\right) \right\rangle =\Psi \left( \frac{\mathbf{r}}{l_1},\frac{l_2}{
l_1},\frac{l_3}{l_1},\frac{l_4}{l_1}\right) .
\]

This formula has too many arguments. We define a simpler field that contains
the same information as $\psi \left( \mathbf{x},l_1,l_2\right) .$

From the definition of $\psi $, one has the identity
\[
\psi \left( \mathbf{x},l,l_2\right) =\psi \left( \mathbf{x},l,l_1\right)
\psi \left( \mathbf{x},l_1,l_2\right).
\]
Let us consider the limit $l_2\rightarrow l_1=l.$ Expanding both sides of
the identity in $\delta l=l_2-l_1$, we have
\[
\psi \left( \mathbf{x},l,l_1\right) +\frac{\partial \psi \left( \mathbf{x}
,l,l_1\right) }{\partial l_1}\delta l+\ldots =\psi \left( \mathbf{x}
,l,l_1\right) \left[ 1+\frac{\partial \psi \left( \mathbf{x},l_1,l_2\right)
}{\partial \ln l_2}|_{l_2=l_1}\frac{\delta l}{l_1}+\ldots \right] ,
\]
and
\begin{equation}
\frac{\partial \psi \left( \mathbf{x},l,l_1\right) }{\partial \ln
l_1} =\varphi \left( \mathbf{x},l_1\right) \psi \left(
\mathbf{x},l,l_1\right) , \label{psi}
\end{equation}
where $\varphi \left( \mathbf{x},l_1\right) =\partial \psi \left(
\mathbf{\ x },l_1,l_2\right) /\partial \ln l_2|_{l_2=l_1}$. The
dimesionless field $ \varphi $ is obtained from $\psi \left(
\mathbf{x},l_1,l_2\right) $ through infinitesimal displacement of
$l_2$. Thus, its correlation functions are scale invariant. If the
statistics of $\varphi $ were known, the correlations of $\psi $
might be obtained from the Eq. (\ref{psi}). From ( \ref{psi}),
(\ref{ratio})
\begin{equation}
\varepsilon (\mathbf{x},l)=\varepsilon (\mathbf{x},L)\exp \left[
-\int\limits_l^L\varphi (\mathbf{x},l_1)\frac{dl_1}{l_1}\right] ,
\label{eps}
\end{equation}
This equation determines the smoothed dissipation in terms of the scale
invariant field $\varphi $.

\subsection{Gaussian Self- Similar Model}

Let $L_0$ be the largest scale in the inertial range. We suppose
that $L$ in the Eq. (\ref{eps}) is larger than $L_0$ so that
$\varepsilon (\mathbf{x} ,L)=<\varepsilon >=\varepsilon _0$ has a
constant value. In this subsection, $\varphi (\mathbf{x},l)$ is
supposed to be a Gaussian field. Gaussian fields are determined by
the correlation functions of first and second order. The mean
value $\overline{\varphi }=\left\langle \varphi (\mathbf{x}
,l)\right\rangle $ is some constant in the inertial range as it
follows from spatial homogeneity and scale invariance of $\varphi
.$
\[
\left\langle \varepsilon (\mathbf{x},l)\right\rangle =\varepsilon _0.
\]
Let us take into account the conservation of the mean flow of energy along
the scale axis. This leads to an additional relation that will be used
later.
\begin{equation}
1=\left\langle \exp \left[ -\int\limits_l^L\varphi
(\mathbf{x},l_1)\frac{dl_1 }{ l_1}\right] \right\rangle.
\label{e1}
\end{equation}
For any Gaussian field $f(l)$ one has \cite{MY75_1}
\begin{eqnarray}
&&\left\langle \exp \left[ -i\int\limits_l^L\theta \left( l_1\right) f\left(
l_1\right) dl_1\right] \right\rangle  \label{my} \\
&=&\exp \left[ -i\int\limits_l^L\theta \left( l_1\right) \left\langle
f\left( l_1\right) \right\rangle dl_1-\frac
12\int\limits_l^Ldl_1\int\limits_l^Ldl_1\theta \left( l_1\right) \theta
\left( l_2\right) \left\langle f\left( l_1\right) f\left( l_2\right)
\right\rangle \right].  \nonumber
\end{eqnarray}
If we choose here $f\left( l\right) =\varphi (\mathbf{x},l),\theta \left(
l\right) =-i/l$ , then from (\ref{e1})
\[
\ \exp \left[ -\int\limits_l^L\left\langle \varphi (\mathbf{x}
,l_1)\right\rangle \frac{dl_1}{l_1}+\frac 12\int\limits_l^L\frac{dl_1}{l_1}
\int\limits_l^L\frac{dl_2}{l_2}\left\langle \varphi (\mathbf{x},l_1)\varphi
( \mathbf{x},l_2)\right\rangle \right] =1,
\]
and
\begin{equation}
\int\limits_l^L\frac{dl_1}{l_1}\int\limits_l^L\frac{dl_2}{l_2}\left\langle
\varphi (\mathbf{x},l_1)\varphi (\mathbf{x},l_2)\right\rangle
=2\int\limits_l^L\left\langle \varphi
(\mathbf{x},l_1)\right\rangle \frac{ dl_1}{ l_1}.  \label{mu}
\end{equation}

Formulae (\ref{eps}), (\ref{mu}) give
\begin{equation}
\left\langle \ln ^2\varepsilon (\mathbf{x},l)\right\rangle =\ln
^2\varepsilon _0+2\int\limits_l^L\left\langle \varphi (\mathbf{x}
,l_1)\right\rangle \frac{dl_1 }{l_1}.  \label{lne}
\end{equation}
Let us divide the integration interval into $(l,L_0)$ and
$(L_0,L)$. In the first (inertial) range $\left\langle \varphi
(\mathbf{x},l_1)\right\rangle $ is some constant
$\overline{\varphi }$ as it was noticed above. The formula (
\ref{lne}) is rewritten as

\begin{equation}
\left\langle \ln ^2\varepsilon (\mathbf{x},l)\right\rangle
=A+2\overline{ \varphi } \ln \frac{L_0}l,  \label{lne2}
\end{equation}
where $A$ origins from the contribution $A=\ln ^2\varepsilon
_0+2\int\limits_{L_0}^L\left\langle \varphi (\mathbf{x},l_1)\right\rangle
\frac{ dl_1}{l_1}$ that is not self-similar.

The similar formula had been derived using the log-normal model \cite{MY75_2}
\[
\left\langle \ln ^2\varepsilon (\mathbf{x},l)\right\rangle
=A(\mathbf{x} )+\mu \ln \frac Ll .
\]
Parameter $\mu $ is known to be equal $0.2\div 0.4.$ We see that it is
universal and $\overline{\varphi }=\mu /2$.

In order to evaluate from (\ref{eps}) the spatial correlations of higher
order, one needs the second moments of the Gaussian field $\varphi $.
Translation, rotation and scale symmetries give for the pair correlations
\begin{equation}
\left\langle \varphi (\mathbf{x},l_1)\varphi (\mathbf{x+r},l_2)\right\rangle
=\Phi \left( \frac r{l_1},\frac{l_2}{l_1}\right).  \label{f2}
\end{equation}
The correlation function is determined by the two dimensionless factors. In
order to obtain an informative result, one needs to reduce the number of
those factors. In the next section, that problem is solved with the help of
the conformal invariance.

\section{Simple and intermittent conformal invariance}

Besides translations, rotations and the scale transformations, the conformal
group in 3 dimensions includes the inversion about the unit circle

\begin{equation}
x_i^{\prime }=x_i/x^2,  \label{c12}
\end{equation}
(for details see, for example, \cite{MS69}). The conformal group is the
simplest extension of the scale one.

To see this, let us consider the transformation of an
infinitesimal displacement under the inversion. Let
$\mathbf{y}=\mathbf{x}+\delta \mathbf{ x }$, where $\delta
\mathbf{x}$ is an infinitesimal displacement. The inversion
transforms the vector $\delta \mathbf{x}$ according to

\begin{equation}
\delta x_i^{\prime }=\frac 1{x^2}\left( \delta _{ij}-2\frac{x_ix_j}{x^2}
\right) \delta x_j.  \label{13}
\end{equation}
This transformation is the rotation by the orthogonal matrix

\begin{equation}
\Delta _{ij}(\vec x)=\delta _{ij}-2\frac{x_ix_j}{x^2}.  \label{14}
\end{equation}
and the dilatation in $1/x^2$ times. Therefore, the special conformal
transformation (\ref{c12}) locally looks as a combination of rotation and
dilatation.

The simple scaling is often complemented by the conformal invariance which
is interpreted as the local scale invariance \cite{pol70}. It has been
proved mathematically that for a certain class of the Lagrangian field
theories the conformal invariance follows from the scale one \cite{MS69},
\cite{GW70}.

\subsection{Conformal theory for dimensionless fields}

We defined the smoothed fields as integrals over the sphere of radius $l$.
The scale transformation transforms $l$ into $Kl.$ While the conformal
transformation, the dilatation factor $K$ depends on the point. To find this
dependance, let us consider the conformal transformation of the sphere
\[
\left( \mathbf{x}-\mathbf{a}\right) ^2=l^2.
\]

After the special conformal transformation the center of the sphere and its
radius become
\begin{equation}
\mathbf{a}_1=\frac{\mathbf{a}}{a^2-l^2},  \label{a}
\end{equation}
\begin{equation}
l_1=\frac{l}{\left| a^2-l^2\right| }.  \label{ll}
\end{equation}
We suppose that the correlation functions of $\varphi $ are invariant to the
transformations (\ref{a}), (\ref{ll}). The scale $l$ transforms like an
additional imaginary coordinate. In this respect, our conformal
transformation (\ref{a}), (\ref{ll}) is similar to that in the relativistic
field theories.

We suppose that the correlation functions of $\varphi (\mathbf{x},l)$ are
invariant to the above transformations. Translation, rotation and scaling
symmetries has led to (\ref{f2}). The conformal invariance imposes the
additional restriction. The function $\Phi $ may depend on the single
parameter $\left( l_1^2+l_2^2-r^2\right) /l_1l_2$. This parameter can be
checked to be invariant to the conformal (\ref{a}), (\ref{ll}) and to other
above transformations. Therefore, the pair correlation have to be of the
form
\begin{equation}
\left\langle \varphi (\mathbf{x},l_1)\varphi (\mathbf{x+r},l_2)\right\rangle
=\Phi \left( \frac{l_1^2+l_2^2-r^2}{l_1l_2}\right) .  \label{phi}
\end{equation}

\subsection{Log-normal conformal theory for spatial correlations}

In this subsection we suppose that $\varphi (\mathbf{x},l)$ is not only
invariant to conformal transformations but is Gaussian distributed also.
Formula (\ref{eps}) gives

\[
\left\langle \varepsilon (\mathbf{x},l)\varepsilon (\mathbf{x+r}
,l)\right\rangle =\ \varepsilon _0^2\left\langle \exp \left[
-\int\limits_l^L\left[ \varphi (\mathbf{x},l_1)+\varphi
(\mathbf{x+r} ,l_1)\right] \frac{dl_1}{l_1}\right] \right\rangle .
\]
With the help of (\ref{my}), the mean of the exponent is written as the
exponent of mean value of an expression. Using the Eq. (\ref{mu}), we have
\begin{eqnarray*}
&&\ \left\langle \varepsilon (\mathbf{x},l)\varepsilon
(\mathbf{x+r} ,l)\right\rangle  \\ &=&\ \ \varepsilon _0^2\exp
\left\{ -2\overline{\varphi }\ln \frac
Ll+\int\limits_l^L\frac{dl_1}{l_1}\int\limits_l^L\frac{dl_2}{l_2}\left[
\begin{array}{c}
\left\langle \varphi (\mathbf{x},l_1)\varphi (\mathbf{x},l_2)\right\rangle
\\
+\left\langle \varphi (\mathbf{x},l_1)\varphi (\mathbf{x+r}
,l_2)\right\rangle
\end{array}
\right] \right\}  \\ &=&\ \varepsilon _0^2\exp \left\{
\int\limits_l^L\frac{dl_1}{l_1}
\int\limits_l^L\frac{dl_2}{l_2}\left\langle \varphi
(\mathbf{x},l_1)\varphi ( \mathbf{x+r},l_2)\right\rangle \right\}
.
\end{eqnarray*}

The last integrals is analyzed in polar coordinates $\lambda ,\chi
$ in $ l_1,l_2$ space: $\lambda =\sqrt{l_1^2+l_2^2}$, $\sin \chi
=l_2/\sqrt{ l_1^2+l_2^2}$. The region of integration is divided in
the 3 sub-regions 1) $ l\leq \lambda $ $\leq r$, 2) $r\leq \lambda
\leq L_0$, 3) $L_0\leq \lambda \leq L$ .

The last region gives some non-universal contribution $\alpha _3$. In that
region the distance r$\ll \lambda $ and may be omitted. Thus, the
dimensionless contribution $\alpha _3$ is determined by the large scale
structure and does not depend on $r$.

Sub-regions 1 and 2 belong to the inertial range. The scale and conformal
invariance give in the polar coordinates

\begin{equation}
\ \ \left\langle \varepsilon (\mathbf{x},l)\varepsilon (\mathbf{x+r}
,l)\right\rangle =\ \ \varepsilon _0^2\exp \left\{ \alpha
_3+2\int\limits_l^{L_0}\frac{d\lambda }\lambda \int\limits_0^{\frac \pi 2}
\frac{d\chi }{\sin 2\chi }\Phi \left[ \frac 4{\sin ^22\chi }\left( \frac{
\lambda ^2-r^2}{\lambda ^2}\right) ^2\right] \right\}.  \label{e2}
\end{equation}

In the sub-region 2 the main logarithmical divergent term is extracted. In
the remainder convergent contribution the upper limit is replaced by $\infty
$. That approximation gives an error of the order of $O(r^2/L^2)$.

The result of the integration is
\[
\left\langle \varepsilon (\mathbf{x},l)\varepsilon (\mathbf{x+r}
,l)\right\rangle \approx C\varepsilon _0^2\left(
\frac{L_0}r\right) ^\mu ,
\]
\begin{equation}
\ C=\exp \left[ \sum\limits_{i=1}^3\alpha _i(A)\right] ,
\end{equation}
where $\alpha _i,i=1,2,3$ are determined by the integrals over the
subregions 1,2,3.
\begin{eqnarray*}
\alpha _1\left( A\right)  &=&2\int\limits_{l/r}^1\frac{ds}
s\int\limits_0^{\frac \pi 2}\frac{d\chi }{\sin 2\chi }\Phi \left[
4\frac{ s^2-1}{s^2\sin ^22\chi }\right]  \\ &\approx
&2\int\limits_0^1\frac{ds}s\int \frac{d\chi }{\sin 2\chi }\Phi
\left[ 4\frac{s^2-1}{s^2\sin ^22\chi }\right] =const,
\end{eqnarray*}
\begin{equation}
\ \alpha _2(A)=2\int\limits_1^\infty
\frac{ds}s\int\limits_0^{\frac \pi 2} \frac{d\chi }{\sin 2\chi
}\left[ \Phi \left( 4\frac{s^2-1}{s^2\sin ^22\chi } \right) -\Phi
\left( \frac{-4}{\sin ^22\chi }\right) \right] =const, \nonumber
\end{equation}

\[
\ \ \alpha _3(A)=\ =4\int\limits_{L_0}^L\frac{d\lambda }\lambda
\int\limits_0^{\frac \pi 4}\frac{d\chi }{\sin 2\chi }\left\langle \varphi (
\mathbf{x},\lambda \cos \chi )\varphi (\mathbf{x+r},\lambda \sin \chi
)\right\rangle .
\]

\subsection{Spatial correlations of higher order}

Formula (\ref{eps}) gives for the correlation function of $n$th order

\begin{equation}
\left\langle \prod\limits_{i=1}^n\varepsilon (\mathbf{x}_i,l)\right\rangle
=\varepsilon _0^n\left\langle \exp \left[
-\int\limits_l^L\sum\limits_{i=1}^n\varphi (\mathbf{x}_i,l_1)\frac{dl_1}{l_1}
\right] \right\rangle.  \label{en}
\end{equation}
With the help of the Eq. (\ref{my}), the mean of the exponent is written as
the exponent of mean value of an expression. Using the Eq. (\ref{e1}), we
have
\begin{eqnarray*}
&&\left\langle \exp \left[
-\int\limits_l^L\sum\limits_{i=1}^n\varphi ( \mathbf{x
}_i,l_1)\frac{dl_1}{l_1}\right] \right\rangle \  \\ &=&\ \exp
\left\{ -n\overline{\varphi }\ln \frac Ll+\frac 12\int\limits_l^L
\frac{dl_1}{l_1}\int\limits_l^L\frac{dl_2}{l_2}\sum\limits_{i=1}^n\sum
\limits_{j=1}^n\left\langle \varphi (\mathbf{x}_i,l_1)\varphi
(\mathbf{x} _j,l_2)\right\rangle \right\} \\ \ &=&\frac
12\int\limits_l^L\frac{dl_1}{l_1}\int\limits_l^L\frac{dl_2}{l_2}
\sum\limits_{i\neq j}^n\Phi \left[
\frac{l_1^2+l_2^2-r_{ij}^2}{l_1l_2} \right],
\end{eqnarray*}
where $r_{ij}^2=\left( \mathbf{x}_i-\mathbf{x}_j\right) ^2$.

The integral is of the same kind as considered above. Similar
straightforward algebra leads to
\[
\left\langle \prod\limits_{i=1}^n\varepsilon (\mathbf{x}_i,l)\right\rangle
=C^n\varepsilon _0^n\prod\limits_{i\neq j}\left( \frac{L_0}{r_{ij}}\right)
^\mu.
\]

\subsection{Non-Gaussian conformal fields of higher order}

Let us consider the correlation function of $\varphi $ of the 3rd order
\[
\Phi _3\left(
\mathbf{x}_1,l_1,\mathbf{x}_2,l_2,\mathbf{x}_3,l_3,\right)
=\left\langle \varphi \left( \mathbf{x}_1,l_1\right) \varphi
\left( \mathbf{x } _2,l_2\right) \varphi \left(
\mathbf{x}_3,l_3\right) \right\rangle.
\]
Symmetry to translations, rotations and to the scale transformations lead to
the form
\[
\Phi _3\left( \mathbf{x}_1,l_1,\mathbf{x}_2,l_2,\mathbf{x}_3,l_3,\right)
=\Phi ^{\prime }\left( \frac{x_{12}}{l_1},\frac{x_{13}}{l_1},\frac{x_{23}}{
l_1},\frac{l_2}{l_1},\frac{l_3}{l_1}\right),
\]
where $\Phi ^{\prime }$ is some new universal function. There are three
conformal invariants of the same kind as in the Eq. (\ref{phi}). The
conformal symmetric function have to be
\[
\Phi _3\left( \mathbf{x}_1,l_1,\mathbf{x}_2,l_2,\mathbf{x}_3,l_3,\right)
=\Phi ^{\prime \prime }\left( \frac{l_1^2+l_2^2-x_{12}^2}{l_1l_2},\frac{
l_1^2+l_3^2-x_{13}^2}{l_1l_3},\frac{l_3^2+l_2^2-x_{23}^2}{l_3l_2}\right).
\]
The generalization to the correlations of more high order is obvious. The
correlation function have to depend on all independent conformal invariants.

\section{Conclusions}

We started from the modified Kolmogorov theory in terms of the ratios of
smoothed fields. The scale symmetry determines the correlation functions of
those fields as universal functions of dimensionless arguments. The
differential equation (\ref{psi}) expresses the usual fields in terms on the
scale invariant ones. Conformal symmetry diminished the number of
dimensional arguments. For the Gaussian $\varphi $, the formulae of the
log-normal model follows with definite expressions for its parameters in
terms of the integrals of correlations of the conformal field $\varphi $.
Experimental measuring of the correlations of $\varphi $ seems to be
necessary to check the proposed conformal symmetry. The generalization of
the present theory to the vector fields is possible and will be considered
in a separate paper.

\section{Acknowledgments}

The support of the Russian Foundation of Basic Research, Grant No
98-01-00681, is acknowledged. The work was also sponsored in the
frame of the project No 274 of Federal Program of the Integration
of High Education and Basic Research. \bibliographystyle{unsrt}
\bibliography{bib}

\end{document}